\documentclass[12pt,fleqn]{article}
\usepackage{epsfig}
\usepackage{graphicx}
\usepackage{amsmath}
\usepackage{cancel}
\usepackage{scrpage2}

\usepackage[small]{caption}
\usepackage[usenames,dvipsnames]{color}
     \definecolor{hgreen}{rgb}{0,.3,0}
     \definecolor{hred}{rgb}{.3,0,0}
     \definecolor{hblue}{rgb}{0,0,.3}
     \definecolor{LightGray}{gray}{0.95}  

\usepackage[colorlinks=true,
            linkcolor=hblue,
            citecolor=hgreen,
            filecolor=hblue,
            urlcolor=hred]{hyperref} 

\newcommand{\MS}{\overline{\text{MS}}}

\newcommand\pubnumber{TUM-HEP 789/11}
\newcommand\pubdate{January 14, 2011}

\def\napoli{
Institute for Advanced Study, Technische Universit\"at M\"unchen, \\
Lichtenbergstra\ss{}e 2a, D-85748 Garching, Germany\\[1em]
Physik-Department, Technische Universit\"at M\"unchen,\\
James-Franck-Stra\ss{}e, D-85748 Garching, Germany
}

\long\def\symbolfootnote[#1]#2{\begingroup%
\def\thefootnote{\fnsymbol{footnote}}\footnote[#1]{#2}\endgroup} 

\def\mail{\symbolfootnote[1]{estamou@ph.tum.de}}

\def\Title#1{\begin{center} {\Large #1 } \end{center}}
\def\Author#1{\begin{center}{ \sc #1} \end{center}}
\def\Address#1{\begin{center}{ \it #1} \end{center}}

\newcommand\pubblock{\rightline{\begin{tabular}{l} \pubnumber\\
         \pubdate  \end{tabular}}}
\newenvironment{Abstract}{\begin{quotation}  }{\end{quotation}}
\newenvironment{Presented}{\begin{quotation} \begin{center} 
             PRESENTED AT\end{center}\bigskip 
      \begin{center}\begin{large}}{\end{large}\end{center} \end{quotation}}

\begin{document}
\begin{titlepage}
\pubblock

\vfill
\Title{Rare Kaon Decays}
\vfill
\Author{Emmanuel Stamou\mail}
\Address{\napoli}
\vfill

\begin{Abstract}
  Rare Kaon decays provide sensitive probes of short distance physics.
  Among the most promising modes are the $K^+\rightarrow\pi^+\nu\bar\nu $ and
  $K_L\rightarrow\pi^0\nu\bar\nu$ modes due to their theoretical cleanness
  and the dedicated experiments, NA62 and KOTO. The key to the
  success of New Physics searches in these modes is the precision in the standard
  model prediction. We review the status and the recent progress of their
  standard model prediction, and also discuss the predictions and
  New Physics sensitivities of the $K_L\rightarrow\pi^0\ell^+\ell^-$ and
  $K_L\rightarrow\mu^+\mu^-$ decays.
\end{Abstract}

\vfill

\begin{Presented}
CKM 2010\\
the $6^{th}$ International Workshop on the CKM Unitarity Triangle\\
University of Warwick, UK, 6-10 September 2010
\end{Presented}
\vfill
\end{titlepage}
\def\thefootnote{\fnsymbol{footnote}}
\setcounter{footnote}{0}

\section{Introduction}
  Within the standard model (SM), short distance (SD) physics accounts for approximately
  $90\%$, $40\%$ and $40\%$ 
  of the branching fractions of the rare Kaon decays $K\rightarrow\pi\nu\bar\nu$, 
  $K_L\rightarrow \pi^0\ell^+\ell^-$ and $K_L\rightarrow \mu^+\mu^-$, respectively. These
  decays are all governed by a flavour changing neutral current (FCNC) transition and
  therefore loop induced within the SM.
  Loop induced processes are special in the sense that heavy new particles
  may contribute to the decay amplitudes on the same footing as SM particles.
  Thus, they represent ideal probes
  for searching and disentangling New Physics (NP). To succeed in isolating NP
  contributions it is essential to have an accurate SM prediction with good
  theoretical control over both short- and long-distance (LD) effects.
  
  In the following we briefly discuss the general structure of rare Kaon decays
  and to which extend each decay channel can reveal information about degrees of
  freedom beyond the SM. We review the SM prediction of the above-mentioned
  modes and also present the recent electroweak (EW) calculation on the top-quark
  contribution to the two $K\rightarrow \pi\nu\bar\nu$ decays.

\section{Structure of Rare Kaon Decays}
\begin{figure}[]
  \begin{center}
    \includegraphics[scale=0.55]{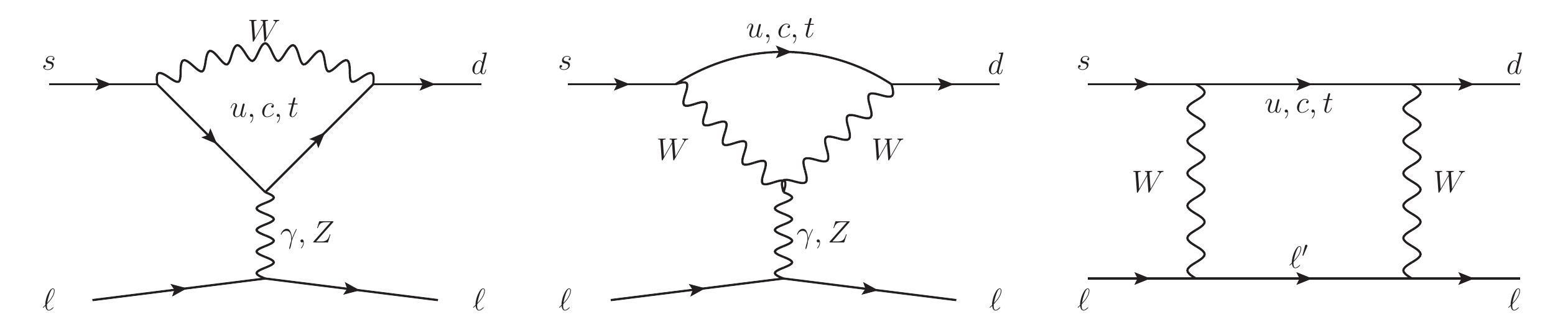}
  \end{center}
  \caption{$\gamma, Z$ penguins and W boxes contributing to the $s\rightarrow d$
  transition at the one-loop level. In the case $l\equiv \nu$ no $\gamma$-penguin
  contributes.}
  \label{fig:boxandpen}
\end{figure}

  Rare Kaon decays proceed within the SM via the quark-level transition $s~\rightarrow~d$ through
  the box and $\gamma,Z$-penguin diagrams of Fig.~\ref{fig:boxandpen}. However, at the
  Kaons' mass scale of approximately $500$~MeV the physics is best described by an appropriate
  effective field theory (EFT) in which heavy particles like top-quark, W- and Z-bosons are
  no longer dynamical degrees of freedom. A master amplitude
  then applies to all meson decays~\cite{Buchalla:1995vs}:
  \begin{equation}
    \mathcal{A}_{\text{decay}}= \sum_i\; \langle Q_i\rangle\; V_{\text{CKM}}^i\; F_i.
    \label{eq:masteramplitude}
  \end{equation}
  $\langle Q_i\rangle$ are matrix elements of effective operators, that cannot be calculated
  perturbatively,  $V_{\text{CKM}}^i$ comprise products of CKM matrix elements and
  $F_i$ are the so-called Wilson coefficients, effective coupling constants that
  parametrise effects of SD physics. The sum in  Eq.~\eqref{eq:masteramplitude}
  extends over all possible operators $Q_i$ generated by the SM or a given NP model and similarly
  over contributions from different up-type quarks in the loops of Fig.~\ref{fig:boxandpen}.
  The EFT approach allows to separate SD from LD physics and therefore to treat
  these contributions independently.
  While the matrix elements of semileptonic decays can be extracted from experimental data,
  the Wilson coefficients depend on the high-energy model and can be calculated
  perturbatively within the SM.
  
  Two points should be stressed regarding 
  Eq.~\eqref{eq:masteramplitude}.
  First, $F_i$ are process independent universal loop functions within the SM,
  meaning that they are the same for B-, D- and K-meson decays. The resulting correlation
  of observables in the SM tests the universality of the loop functions in a model
  independent way (see talk by D.~Straub \cite{Straub:2010ih}).
  Secondly, NP can not only change the Wilson coefficients, but can also generate
  additional operators 
  contributing to the decay amplitude. This would signal a departure from Minimal 
  Flavour Violation \cite{Buras:2000qz}. Both NP effects can affect the branching
  ratios of rare Kaon decays.
  
  The master amplitude of Eq.~\eqref{eq:masteramplitude} also incorporates the
  relative magnitude of each contribution. Its size depends on both the CKM elements,
  conveniently parametrised in terms of $\lambda=\vert V_{us}\vert\approx 0.22$,
  and on the loop functions $F_i$'s. Within the SM, low-energy contributions are suppressed
  with respect to the top contribution due to the Glashow-Iliopoulos-Maiani (GIM)
  mechanism \cite{Glashow:1970gm}. When SD photon penguins contribute, this suppression is
  only logarithmic, but  in their absence (e.g. in the $K\rightarrow\pi\nu\bar\nu$ decays)
  the loop functions are proportional to $m_{\text{top}}^2$, $m_{\text{charm}}^2$, or $\Lambda_{\text{QCD}}^2$
  for the up-quark, and a hard quadratic GIM is at play.
  
  In addition, the CKM factor of the top-quark contribution 
  $\lambda_t=V_{ts}^*V_{td}$ is proportional to $\lambda^5$ and therefore 
  highly suppressed. This strong suppression, absent in B and D decays, can imply a
  significant charm-contribution depending on the mode, but also renders
  rare Kaon decays sensitive to even very small deviations from the CKM picture 
  of $CP$ violation.


\section{\texorpdfstring{$\boldsymbol{K^+\rightarrow \pi^+\nu\bar\nu\; \text{and}\; K_L\rightarrow \pi^0\nu\bar\nu}$}
                        {K+ --> pi+ nu nu and K(L) --> pi0 nu nu}}
  \begin{figure}[t]
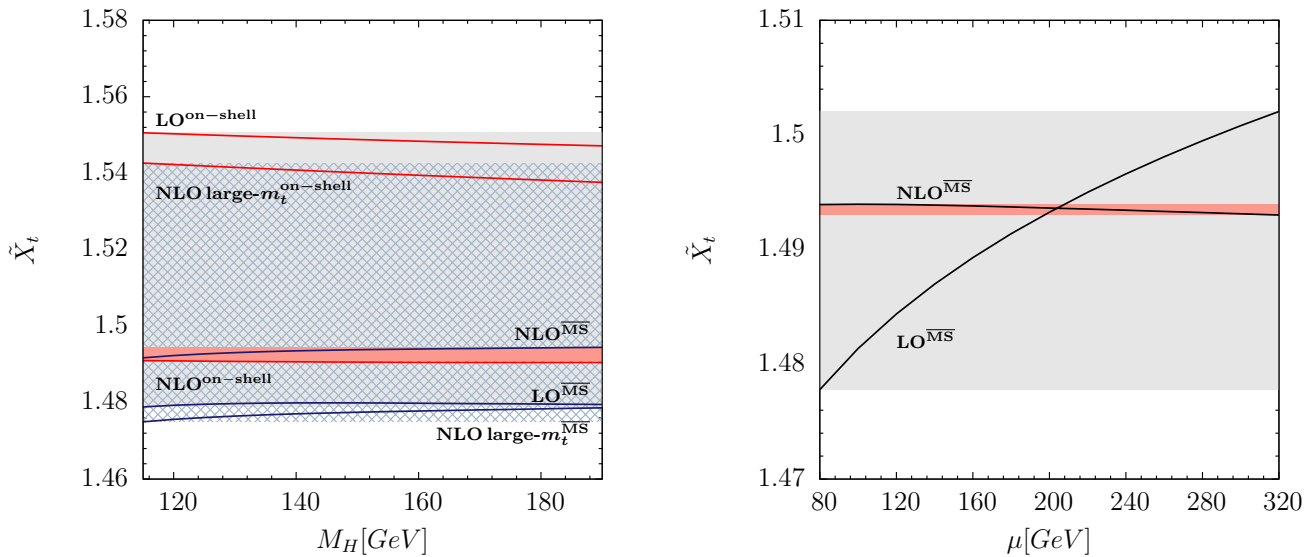

    \hspace*{-18ex}
    \begin{minipage}[t]{0.5\textwidth}
      \scalebox{0.85}{\input{./X_mh_NLO_bandT}}
    \end{minipage}
    \qquad\quad
    \begin{minipage}[t]{0.5\textwidth}
      \scalebox{0.85}{\input{./X_mu_bandT}}
    \end{minipage}
    \caption{Left: comparison of the top-quark Wilson coefficient in the $\MS$ and
    on-shell schemes at LO, NLO large-$m_t$ and NLO. The grey solid band, hashed
    band, and solid red band represent the estimated uncertainties of LO,
    NLO large-$m_t$, and NLO result as a function of the Higgs mass, respectively.
    Right: the uncertainty due to EW renormalisation scale at LO
    and NLO. Again the grey and red band represent the estimated uncertainty before and after 
    the NLO calculation, respectively.}
    \label{fig:XtMSOS}
  \end{figure}
  The weak SM effective Hamiltonian for the two $K\rightarrow\pi\nu\bar\nu$ decays
  reads~\cite{Buchalla:1998ba}
  \begin{equation}
    \mathcal{H}_{\text{eff}}=\frac{4
    G_F}{\sqrt{2}}\frac{\alpha}{2\pi\sin^2\theta_W} \sum_{l=e,\mu,\tau}\left(
    \lambda_c X^{l} + \lambda_t X_t\right) (\bar s_L\gamma_{\mu}d_L)(\bar
    \nu_{lL}\gamma^{\mu}\nu_{lL}) + h.c.
    \label{eq:effectiveHamilKpinunu}
  \end{equation}
  and involves to a good approximation only one effective operator below $\mu=\mu_{c}$ .
  
  The LD part of both decays is known with high precision. The
  matrix elements are extracted from $K_{\ell3}$ decays including isospin-breaking
  ($\kappa_{\nu}^{+,L}$) and long-distance QED effects \cite{Mescia:2007kn}, whereas
  also effects from light quarks and higher dimensional operators ($\delta P_{c,u}$)
  have been estimated using chiral perturbation theory (ChPT) \cite{Isidori:2005xm}.
  
  Regarding the small uncertainties from LD effects, higher order corrections on
  the SD parts become essential for a precise SM prediction. The SD charm contribution,
  denoted by $P_c=\tfrac{1}{\lambda^4}\left( \tfrac{2}{3}X^e+\tfrac{1}{3} X^{\tau} \right)$,
  is negligible for $K_L\rightarrow\pi^0\nu\bar\nu$, but not
  for $K^+\rightarrow\pi^+\nu\bar\nu$, where it amounts to approximately 30\% of its
  branching ratio. So far $P_c$ is known up to next-to-next-to-leading
  order (NNLO) in QCD \cite{Buras:2006gb} and also the next-to-leading
  order (NLO) EW corrections have been calculated \cite{Brod:2008ss}.
  NLO QCD corrections \cite{Buchalla:1998ba,Misiak:1999yg} are known for the top-quark
  contribution, but until recently the NLO EW corrections were only known in the large-$m_t$ limit 
  \cite{Buchalla:1997kz}. However, these were known to poorly
  approximate the full EW corrections (see Fig.~\ref{fig:XtMSOS}) \cite{Buchalla:1997kz}.
  Therefore, the renormalisation scheme of the EW input parameters 
  $\alpha, \sin \theta_W, M_W, M_t$, appearing at LO in Eq.~\eqref{eq:effectiveHamilKpinunu},
  was not clear and accounted for an approximately $2\%$ uncertainty in $X_t$, which
  scales quadratically in the uncertainty of the branching ratios.
   
  Our recent two-loop calculation of the full NLO EW corrections on $X_t$ \cite{Brod:2010hi}
  substantially reduced the renormalisation scheme uncertainty to approximately  $0.2\%$
  and also removed the remaining scale dependence. The comparison of LO and
  NLO results in the $\MS$ and on-shell scheme is illustrated in Fig.~\ref{fig:XtMSOS} together
  with the remaining scale dependence.
  
  The theoretical predictions and the current experimental values for the
  branching ratios then read:
  \begin{align*}
    \hspace*{-6ex}&
    \text{BR}^{\text{theo}}_{K^+\rightarrow \pi^+\nu\bar\nu}= (7.81^{+0.80}_{-0.71}\pm0.29)\times 10^{-11}
    &
    \hspace*{-2ex}\cite{Brod:2010hi}\hspace*{5ex}
    & 
    \text{BR}^{\text{exp}}_{K^+\rightarrow \pi^+\nu\bar\nu}= (1.73^{+1.15}_{-1.05})\times 10^{-10}
    &
    \hspace*{-2ex}\cite{Artamonov:2008qb}
    &
    \\
    \hspace*{-6ex}&
    \text{BR}^{\text{theo}}_{K_L\rightarrow \pi^0\nu\bar\nu}= (2.43^{+0.40}_{-0.37}\pm 0.06)\times 10^{-11}
    &
    \hspace*{-2ex}\cite{Brod:2010hi}\hspace*{5ex}
    & 
    \text{BR}^{\text{exp}}_{K_L\rightarrow \pi^0\nu\bar\nu} < 6.7\times 10^{-8}
    &
    \hspace*{-2ex}\cite{Ahn:2009gb}
    &
    \label{eq:kpinunuBR}
  \end{align*}
  The first error in the theory prediction is parametric, while the second
  summarises the remaining theory uncertainty. For the charged mode the detailed 
  main parametric contributions are ($V_{cb}: 56\%$, $\bar\rho:
  21\%$, $m_c: 8\%$, $m_t: 6\%$, $\bar\eta: 4\%$, $\alpha_s: 3\%$,
  $\sin^2\theta_W: 1\%$), while the main theoretic contributions are ($\delta
  P_{c,u}: 46\%$, $X_t(\text{QCD}): 24\%$, $P_c: 20\%$, $\kappa_{\nu}^+:
  7\%$, $X_t(\text{EW}): 3\%$), respectively. Similarly for the neutral mode, the parametric
  uncertainties are ($V_{cb}: 54\%$, $\bar\eta: 39\%$, $m_t: 6\%$) and the theoretical are
  ($X_t(\text{QCD}): 73\%$, $\kappa_{\nu}^L: 18\%$, $X_t(\text{EW}): 8\%$,
  $\delta P_{c,u}: 1\%$), respectively.
  
\section{\texorpdfstring{$\boldsymbol{K_L\rightarrow \pi^0\ell^+\ell^-}$}
                        {K(L) --> pi0 l+ l-}}
  \begin{figure}[]
  \begin{center}
    \includegraphics[scale=0.6]{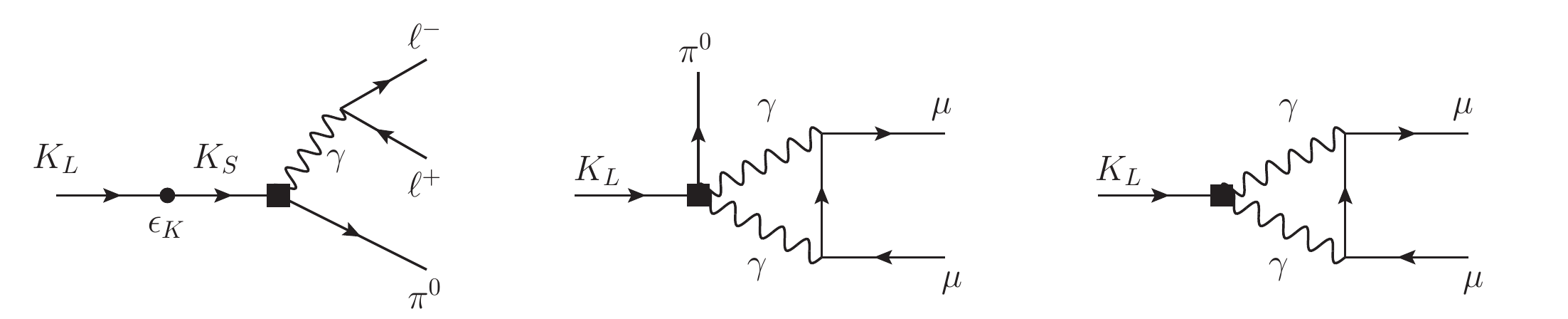}
  \end{center}
  \caption{$\gamma$- and $\gamma\gamma$- penguins contributing to the LD
  part of the $K_L\rightarrow\pi^0\ell^+\ell^-$ and $K_L\rightarrow\mu^+\mu^-$
  branching ratios.
  }
  \label{fig:photons}
  \end{figure}
  In the two $K_L\rightarrow \pi^0\ell^+\ell^-$ modes three
  different contributions compete in magnitude: direct-$CP$-violating (DCPV), indirect-$CP$-violating
  (ICPV) and $CP$-conserving (CPC) contribution. Long distance effects are much larger here. Still, the importance
  of these decay channels lies in their sensitivity to SD contributions from more than one
  operator. Let us now look at the three distinct contributions.
  
  {\bf DCPV:} high-energy top and charm contributions generate both vector, $Q_{7V}=(\bar sd)_V(\bar
  \ell\ell)_V$, and axial-vector, $Q_{7A}=(\bar sd)_V(\bar\ell\ell)_A$, dimension-six
  operators. Their Wilson coefficients are known to NLO QCD \cite{Buras:2000qz}
  and their matrix elements are also extracted with a few per mil precision from
  $K_{\ell3}$ decays \cite{Mescia:2007kn}. One difference between the electronic and the
  muonic channels is the suppression of the electronic axial-vector matrix element
  due to the smallness of the electron mass. 

  {\bf ICPV:} in this case the $K_L$ decays through its small $CP$ even component
  via the $\gamma$-penguin in Fig.~\ref{fig:photons}.
  The size of this smaller component is described by the parameter $\epsilon_K$.
  We can therefore relate this contribution to the decay
  $K_S(\approx K_1)\rightarrow \pi^0\ell^+\ell^-$
  using ChPT. The light meson loops are subleading and the amplitude is 
  dominated by the chiral counterterm $a_S$ \cite{D'Ambrosio:1998yj}, whose $20\%$
  uncertainty completely dominates the error for the $K_L\rightarrow\pi^0 \ell^+\ell^-$
  rates \cite{Mescia:2006jd}.
  Although the absolute  value of $a_S$ has been extracted from NA48 measurements, its
  sign remains unknown. Therefore, it is still an open question whether the ICPV 
  contribution interferes constructively or destructively with the DCPV contribution.
  Apart from a
  more precise $K_S\rightarrow \pi^0\ell^+\ell^-$ measurement, the sign of $a_S$
  can also be determined from the measurement of the integrated forward-backward
  asymmetry $A_{FB}^\ell$ in $K_L\rightarrow \pi^0\mu^+\mu^-$ \cite{Mescia:2006jd}.


  {\bf CPC:} is a purely LD effect described by the $\gamma\gamma$-penguin ChPT
  diagram in Fig.~\ref{fig:photons}. The contribution is helicity suppressed
  for the electronic mode, a further difference between the two modes. For the
  muonic mode the meson loops are finite at $\mathcal{O}(p^4)$, while higher order
  effects of $\mathcal{O}(p^6)$ have been partially estimated from the 
  $K_L\rightarrow\pi^0\gamma\gamma$ ratio, reducing the uncertainty to
  approximately $30\%$ \cite{Isidori:2004rb}.

  The current SM predictions and experimental values for the branching ratios
  for the two decays then read:
  \begin{align*}
    \hspace*{-6ex}&
    \text{BR}^{\text{theo}}_{K_L\rightarrow \pi^0 \mu^+\mu^-}= 1.41^{+0.28}_{-0.26}\;(0.95^{+0.22}_{-0.21})\times 10^{-11}
    &
    \hspace*{-2ex}\cite{Mescia:2006jd}\hspace*{6ex}
    & 
    \text{BR}^{\text{exp}}_{K_L\rightarrow \pi^0\mu^+\mu^-}< 3.8\times 10^{-10}
    &
    \hspace*{-2ex}\cite{AlaviHarati:2000hs}
    &
    \\
    \hspace*{-6ex}&
    \text{BR}^{\text{theo}}_{K_L\rightarrow \pi^0 e^+e^-}= 3.54^{+0.98}_{-0.85}\;\,(1.56^{+0.62}_{-0.49})\times 10^{-11}
    &
    \hspace*{-2ex}\cite{Mescia:2006jd}\hspace*{6ex}
    & 
    \text{BR}^{\text{exp}}_{K_L\rightarrow \pi^0 e^+e^-} <2.8\times 10^{-10}
    &
    \hspace*{-2ex}\cite{AlaviHarati:2003mr}
    &
  \end{align*}
  Theory predictions correspond to constructive interference of DCPV with ICPV
  contribution, while predictions in brackets to destructive interference.

\section{\texorpdfstring{$\boldsymbol{K_L\rightarrow \mu^+\mu^-}$}
                        {K(L) --> mu+ mu-}}
  $K_L\rightarrow \mu^+\mu^-$ is a $CP$ conserving decay, whose SD sensitivity
  is obscured by large LD effects. 
  The top contribution is known up to NLO QCD \cite{Buchalla:1993bv} and the charm
  contribution up to NNLO QCD \cite{Gorbahn:2006bm}. Indirect $CP$ violation is negligible
  in this channel. However, in contrast to the $K_L\rightarrow\pi^0\mu^+\mu^-$ case, the LD
  $\gamma\gamma$-penguin is not finite (Fig.~\ref{fig:photons}). The absorptive part of the
  $\gamma\gamma$-loop, extracted from $K_L\rightarrow\gamma\gamma$, accounts to
  approximately $98\%$ of the measured branching ratio, $\text{BR}^{\text{exp}}_{K_L\rightarrow\mu^+\mu^-}
  =6.84(11)\times 10^{-9}$ \cite{Nakamura:2010zzi}, while dispersive
  part of the $\gamma\gamma$-contribution is divergent in ChPT and can therefore be
  estimated from $K_L\rightarrow \gamma^*\gamma^*$ only with a large uncertainty 
  \cite{Isidori:2003ts}. On top of these difficulties, the $\gamma\gamma$-contribution
  interferes with an unknown sign with the SD contribution, which further
  increases the theoretical uncertainty of the channel. Therefore, the interesting
  SD part of the decay mode cannot be extracted accurately. Nevertheless, useful
  NP constraints can be derived using its measured branching
  ratio.

\section{New Physics and Outlook}
  The theoretical cleanness of the $K\rightarrow\pi\nu\bar\nu$ modes
  promotes both to excellent probes of NP, especially in view of the dedicated
  experiments, NA62 at CERN and KOTO at JPARC, which aim at measuring the charged
  and neutral mode, respectively, with an expected accuracy of 15\%.
  These decays have therefore been studied extensively in models
  beyond the SM. A summary of predictions in different models is presented in
  \cite{MesciaPlot}. This analysis points out the possibility of
  large effects in both decay modes and also illustrates the correlation
  between the predictions of the two decays in a large class of NP models.
  
  NP effects on the branching ratios of the $K_L\rightarrow\pi^0\ell^+\ell^-$ modes
  have also been considered, in a model-independent way. A measurement of the decays
  could test contributions from operators not generated by the SM.
  Effects from scalar, pseudoscalar, tensor, and pseudotensor
  operators, generated by NP with or without helicity suppression, have been
  studied and can be disentangled by a more precise measurement of both modes
  \cite{Mescia:2006jd}.

  To conclude, rare Kaon decays are very clean and sensitive probes of NP. The
  observation of deviations from the SM by the dedicated experiments would be
  a clear NP signal. Also, the use of both predictions and measurements
  as constraints of NP can shed light on NP patterns and the structure of flavour
  violation. 
  
\section*{Acknowledgements}
  \addcontentsline{toc}{section}{Acknowledgements}
  I would like to thank the conveners of working group III and the organisers of CKM2010
  for the interesting time in Warwick, Joachim Brod and Martin Gorbahn for their support
  and the fruitful collaboration, and Andrzej Buras for comments on the manuscript.
  This research was supported
  by the DFG cluster of excellence ''Origin and Structure of the Universe''.

\addcontentsline{toc}{section}{References}
\bibliography{Kdecays}
  \bibliographystyle{./bibstyles/proceedings}

\end{document}